# THE EARLY PHASE OF MULTIPLE PROTO-STELLAR SYSTEM EMERGED FROM COLLAPSE OF MOLECULAR CLOUD UNDER VARIOUS INITIAL THERMAL STATES

Main Author: Rafil Riaz

*Institute of Space & Planetary Astrophysics (ISPA), University of Karachi, 75270, Pakistan*
rafilriaz@yahoo.com

Coauthor (1): Suhail Zaki Farooqui

*Faculty of Engineering Sciences, National University of Science & Technology, Islamabad, PNS - JAUHAR, H. I. Rahamatullah Road, Karachi – 75350, Pakistan*
drsuzaki@hotmail.com

Coauthor (2): Siegfried Vanaverbeke
*KU Leuven Campus Kortrijk, E.-Sabbelaan 53, 8500 Kortrijk, Belgium*
*Centre for mathematical Plasma Astrophysics, Department of Mathematics, KU Leuven, Celestijnenlaan 200B, 3001 Leuven, Belgium*
Sigfried.Vanaverbeke@kuleuven-kortrijk.be

Abstract:

An attempt is made here to revisit structure formation in a proto-stellar cloud during the early phase of evolution. Molecular cloud subjected to a set of various initial conditions in terms of initial temperature and amplitude of azimuthal density perturbation is investigated numerically. Special emphasis remained on the analysis of ring and spiral type instabilities that have shown dependence on certain initial conditions chosen for a rotating solar mass cloud of molecular hydrogen. Generally, a star forming hydrogen gas is considered to be initially at 10K. We have found that a possible oscillation around this typical value can affect the fate of a collapsing cloud in terms of its evolving structural properties leading to proto-star formation. We explored the initial temperature range of cloud between 8K to 12K and compared physical properties of each within the first phase of proto-star formation. We suggest that the spiral structures are more likely to form in strongly perturbed molecular cores that initiate their phase of collapse from temperatures below 10K. Whereas, cores with initial temperatures above 10K develop, instead of spiral, a ring type structure which subsequently experiences the fragmentation. A transition from spiral to ring instability can be observed at a typical initial temperature of 10K.

*ISM: clouds, hydrodynamics stellar dynamics (stars:) binaries: general stars: low-mass, brown dwarfs methods: numerical*

Initial conditions (ICs), Tree gravity code (TGC), Smoothed particle hydrodynamics (SPH),

Self-gravitating fragment (SGF)

# 1. Introduction:

Molecular clouds host the formation of variety of proto-stars among which low-mass proto-stellar systems have ever remained as a class that requires special attention. This is mainly due to their unavailability for any astronomical observation as these systems during their initial phases of evolution hardly radiate and they grow inside a thick envelop of gas and dust that keep their secrets alive. However, for the last two decades, these star forming regions have been analyzed in some details through indirect indicators such as CO and $NH_3$ emission lines observation (Maddalena et al., 1986; Bally et al., 1987). The CO emission line from dark clouds in the Orion (Bally et al. 1987) and Taurus (Ungerechts & Thaddeus 1987) have given indications of filamentary and clumpy structures in the interstellar gas clouds. Similarly, the elongated shapes resembling prolate ellipsoids are found in most of the molecular cloud cores (Myers et al. 1991). This has revealed the structure along with other fundamental properties of these dark regions of interstellar medium. For instance, it has been found that the pre-stellar cores exist in various outer radii, and they differ significantly in their central density and temperatures (Stamatellos et al. 2007). It is assumed that these variations in thermal properties from core to core are dependent on their exposure to radiation field lurking around due to presence of other stars in the vicinity (G. Duchêne et al., 2007). In the present paper we concentrate on a numerical model to investigate the possible effects of such thermal changes that can have implications on the overall evolution of these pre-stellar cores that when undergo self-gravitational collapse may give birth to low-mass multiple proto-stellar systems. Numerical scheme has emerged in the recent past as the most powerful tool to explore the phenomena of star formation. The numerical model can efficiently follow the dynamical collapse of rotating pre-stellar cores that eventually after passing through the free fall time show fragmentation hence giving birth to multiple proto-stellar systems which could be in form of binaries, triplets, quadruples, etc. Numerical studies have also proposed various other formation mechanisms that lead to such systems of low-mass gravitationally bound objects (Whitworth et al., 2007). Since we are interested here in the study of formation of a quadruple system under various possible thermal states, hence we select a set of models of cloud with mode $m = 4$ taking into account the rotationally-driven fragmentation scenario. Although as described above, an $m = 2$ mode suits more when it comes to model the molecular cloud cores due to their elongated shapes but it has been assumed here that along with elongated structures some spheroidal oblate cores may form as well.

# 2. Brief note on the code:

GRADSPH is a parallel SPH code in FORTRAN 90 based on a Tree Code Gravity (TCG) method for evolving originally 3-D self-gravitating astrophysical fluids. Following (Gingold and Monaghan, 1977) the set of key equations of motion and energy of SPH particles are described as

$$\frac{dv_i}{dt} = -\sum_{j=1}^{N} m_j \left(\frac{p_i}{\rho_i^2} + \frac{p_j}{\rho_j^2}\right) \nabla W(r_{ij}, h) \quad \ldots \ldots (1)$$

$$\frac{du_i}{dt} = \frac{p_i}{\rho_i^2} \sum_{j=1}^{N} m_i v_{ij} \nabla_i W(r_{ij}, h) \quad \ldots \ldots (2)$$

An iterative approach based convergence criterion is followed in updating the smoothing length during the evolution. This ensures a constant mass within the smoothing sphere of each particle throughout simulation even under extreme density gradients.

$$\frac{4\pi}{3} (\sigma h_i)^3 \rho_i = m_i N_N = constant \quad \ldots \ldots (3)$$

$$N_N = \frac{4\pi}{3} (\sigma \eta)^3 \quad \ldots \ldots (4)$$

$$\frac{|h_{new} - h_{old}|}{h_{old}} < \varepsilon \quad \ldots \ldots (5)$$

where, $\varepsilon = 10^{-3}$.

The code updates the particles properties on their own time step by utilizing the predictor-corrector scheme. Barnes-Hut tree is used for treating self-gravity and updating the neighbor list of the particles. Cubic spline kernel is used to soften the gravity and the following form for gravitational acceleration of particle '$i$' is used.

$$g_i = -G \sum_{j=1}^{N} \frac{m_j}{2} [\phi(r_{ij}, h_i) + \phi(r_{ij}, h_j)] e_{ij}$$

$$- \frac{G}{2} \sum_{j=1}^{N} \left[\frac{\zeta_i}{\Omega_i} \nabla_i W(r_{ij}, h_i) + \frac{\zeta_i}{\Omega_j} \nabla_i W(r_{ij}, h_j)\right]$$

$$\ldots \ldots (6)$$

where,

$$\Omega_i = 1 - \frac{\partial h_i}{\partial \rho_i} \sum_{j=1}^{N} m_j \frac{\partial W(r_{ij}, h_j)}{\partial h_i} \quad \ldots \ldots (7)$$

$$\zeta_i = \frac{\partial h_i}{\partial \rho_i} \sum_{j=1}^{N} m_j \frac{\partial \phi(r_{ij}, h_i)}{\partial h_i} \quad \ldots \ldots (8)$$

In the above equation $\phi(r_{ij}, h_i)$ represents the softened gravitational potential of a particle and the derivatives of $\phi$ with respect the smoothing length are used in GRADSPH directly from the tabulation given in (Price & Monaghan, 2007). The opening angle for gravity computation is set to $\theta = 0.7$. A set of time-dependent artificial viscosity terms $\alpha = 1.0$ and $\beta = 2.0$ is also used in the code while treating the short-range forces. A further detailed description of the code along with its ability to reproduce certain bench mark cases related to fluid dynamical phenomena that depicts efficiency of the code can be found in (S. Vanaverbeke et al., 2007).

## 3. Models Description:

It is believed that a very low-level turbulence i.e. $E_{turb}/E_{grav} \sim 5\%$ of initial core can result the formation of a multiple system (Commerçon et al., 2008). We in the present effort, however, have considered only the effect of rotation and neglected any possible role of turbulent fragmentation (Kevin L.

Luhman, 2012) and a magnetic field effects (Alan P. Boss et al., 2000) on the development of the system as a whole. This is done merely for the sake of simplicity of our models. We select certain cloud parameters that remain constant in each of the models considered. For instance, mass of the cloud is selected as M = 1$M_\odot$, radius of the cloud is R = 5 x$10^{16}$ cm, rigid body rotation assigned to the core is ω = 7 x $10^{-13}$ rad/s. Cloud chemistry follows mean molecular weight µ=3. Identical uniform initial density of cloud as ρ = 3.8 x $10^{-18}$ g/cm$^3$ is being considered. The two dimensionless parameters α & β representing the ratios of thermal energy to gravitational energy, and rotational energy to gravitational energy; respectively are described as

$$\alpha = \frac{5}{2} \frac{RkT}{GM\mu m_h} \quad \ldots \ldots (9)$$

$$\beta = \frac{1}{3} \left( \frac{R^3 \omega^2}{GM} \right) \quad \ldots \ldots (10)$$

In this present study the ratio of thermal to gravitational energy varies according to the selected values of temperature in models, whereas ratio of rotational to gravitational energy remains constant in each model as β = 0.16. The mean freefall time derived from relation $(3\pi/32G\rho_\circ)^{1/2}$ is $t_{ff}$ = 1.07 x $10^{12}$s.

An azimuthally perturbed density of amplitudes A = 10% and 25% with mode m = 4 are introduced in the uniform cloud. The physical parameters used in models are converted into dimensionless forms following G = M = R = 1. In order to effectively capture the thermal behavior of a gas, we have used the barotropic equation of state. This makes the simulation of rotating cloud more physical. (Tohline, 1982) and (Masunaga & Inustsuka, 2000) described that it is possible to approximate the core without resolving the radiative transfer. Hence, the corresponding pressure of gas and sound velocity are determined as

$$P = \rho\, c\_iso^2 \left[ 1 + \left( \frac{\rho}{\rho_{crit}} \right)^{2/3} \right] \ldots \ldots (11)$$

$$c = c\_iso \left[ 1 + \left( \frac{\rho}{\rho_{crit}} \right)^{\gamma-1} \right]^{1/2} \ldots \ldots (12)$$

Where $\rho_{crit}$ = 5 x $10^{-14}$ g/cm$^3$, c_iso is the isothermal sound speed, and γ = 5/3.

(Bate and Burkert, 1997) showed that to avoid artificial fragmentation during collapse of a cloud, simulations must follow a resolution criterion. This imposes a limit to the minimum number of SPH particles used in a collapse test (B. Commercon et al., 2008). The mass within the smoothing sphere $M_{sph}$ must always be less than the local Jeans mass. The Jeans mass is

$$M_J = \frac{\pi^{3/2} c_{iso}^3}{G^{3/2} \rho^{1/2}} \ldots \ldots (13)$$

and the minimum resolvable mass (mass within the smoothing sphere) is $M_{Res} = 2N_N$ m, where $N_N$ is the neighbor count for every SPH particles here chosen as 50. This is achieved by introducing a new subroutine into the original GRADSPH code that guarantees a constant neighbor count for every SPH particle. Following (Arreaga et al., 2007) the Jeans condition can be related to the maximum mass of SPH particle. This limit for ρ = $\rho_{crit}$ becomes:

$$m < m_{max} \sim \frac{\pi^{\frac{3}{2}} c_{iso}^3}{2 N_N \rho_{crit}^{\frac{1}{2}} G^{\frac{3}{2}}} \quad \ldots\ldots (14)$$

With $M_0 = 1 M_\odot$, the critical number of SPH particles essential to resolve fragmentation can be estimated as $N_p > M_0/m_{max}$. The assumption that once the Jeans criterion is satisfied at critical density it will continue to be so all the time in the entire domain of simulation certainly needs a revisit. We, however, stick to the consideration that as the gas experiences a phase transition from isothermal to adiabatic; the adiabatic Jeans mass behaves as an increasing function of density. Thus the Jeans mass remains above the mass contained within smoothing sphere. Although there exists a decrease in Jeans mass while the collapsing gas remains isothermal but as the phase transition from isothermal to adiabatic regime occurs, the value of Jeans mass starts increasing.

It is important to mention that the assuming initial isothermal state of collapsing cloud till the gas reaching the critical density is not a very good approximation as recent studies have revealed that the temperature in collapsing clouds can vary by as much as a factor of 2 or more above and below the usually assumed constant value of 10K (Larson, 2005). In the present study, we have not adopted this possible variation of temperature during the initial phase of collapse and each cloud considered in our models remained in isothermal state until the adiabatic regime takes over. However, initiating collapse models with a range of constant initial temperatures from 8K to 12K by a difference of 2 tries to consider the factor of such change in a rather crude sense. We try to address the transition from phase of different isothermal states of cloud i.e. (8K → 12K) to an adiabatic state.

In each of our models we used number of SPH particle $N_{SPH} = 250025$ that though gives relatively a medium resolution but yet falls well inside the Jeans mass criterion described above according to which a certain minimum number of SPH particles must be used to rule out any possibility of artificial fragmentation.

To study the characteristic of proto-stellar disk formed as a result of collapse of self-gravitating molecular cloud in each of our models we used Toomre Q-parameter analysis (Toomre, 1964) according to which thin disks are unstable to fragmentation if the Q-parameter takes value smaller than unity, where the parameter itself is defined as

$$Q = \frac{c\kappa}{\pi G \Sigma} \quad \ldots\ldots (15)$$

where, c is the sound speed, $\kappa$ is the epicyclic frequency, and $\Sigma$ is the surface density. For column density plots mentioned in the paper, we used an open sourceSPH data visualization tool 'SPLASH' (Price, 2007).

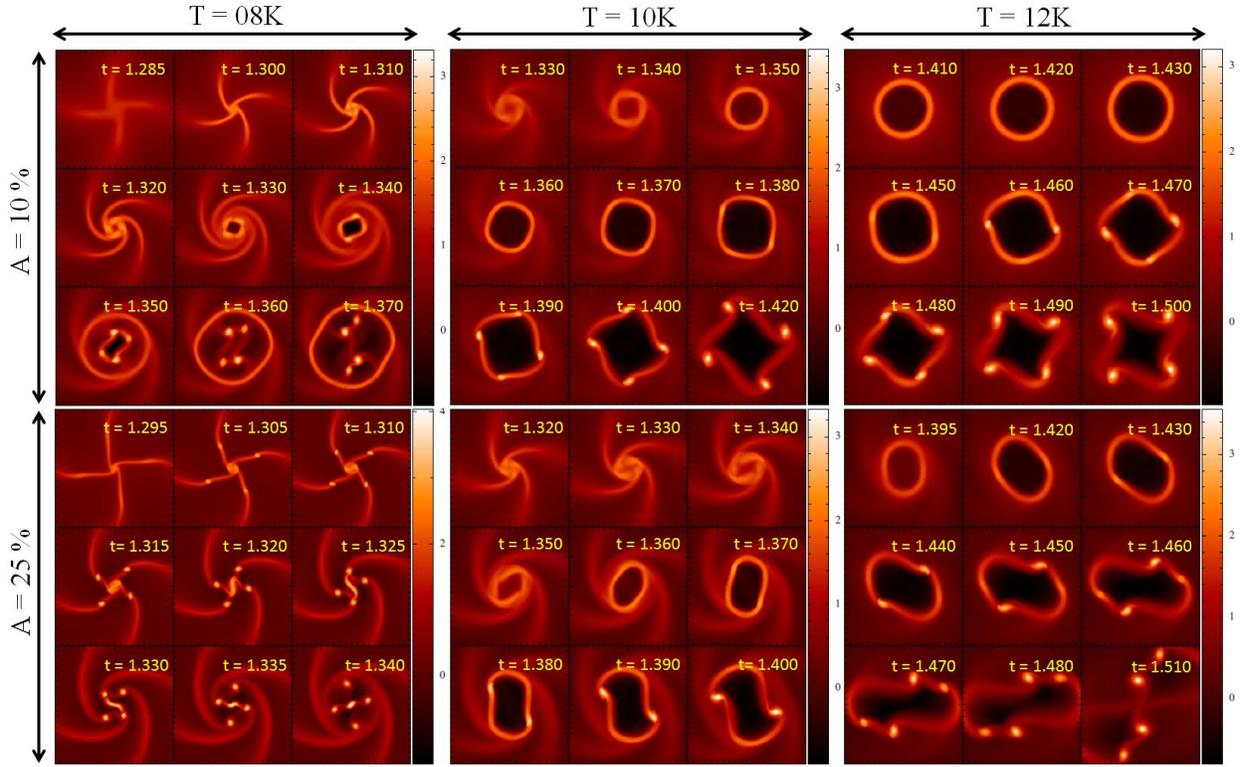

Fig.1- Snapshots of column density viewed parallel to the rotation axis during the evolution of collapse of cloud for models M1, M2, M3 (first row) and for models M4, M5, M6 (second row). The dimensions of each square for xy-plane follow dimensionless units as 0.1 x 0.1. Color coded bar in each panel follows $\log(\rho)$ in dimensionless units. Time in each panel is shown in units of $t/t_{ff}$. Each calculation was performed with 250025 SPH particles.

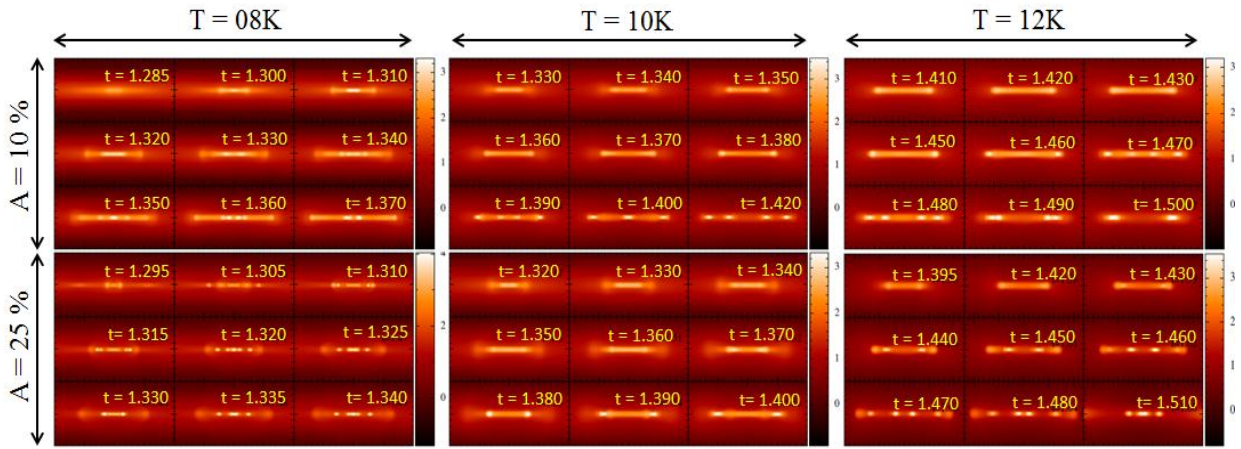

Fig.2 - Snapshots of column density viewed orthogonal to the rotation axis during the evolution of collapse of cloud for models M1, M2, M3 (first row) and for models M4, M5, M6 (second row). The dimensions of each square for xz-plane follow dimensionless units as 0.1 x 0.05. Color coded bar in each panel follows $\log(\rho)$ in dimensionless units. Time in each panel is shown in units of $t/t_{ff}$. Each calculation was performed with 250025 SPH particles.

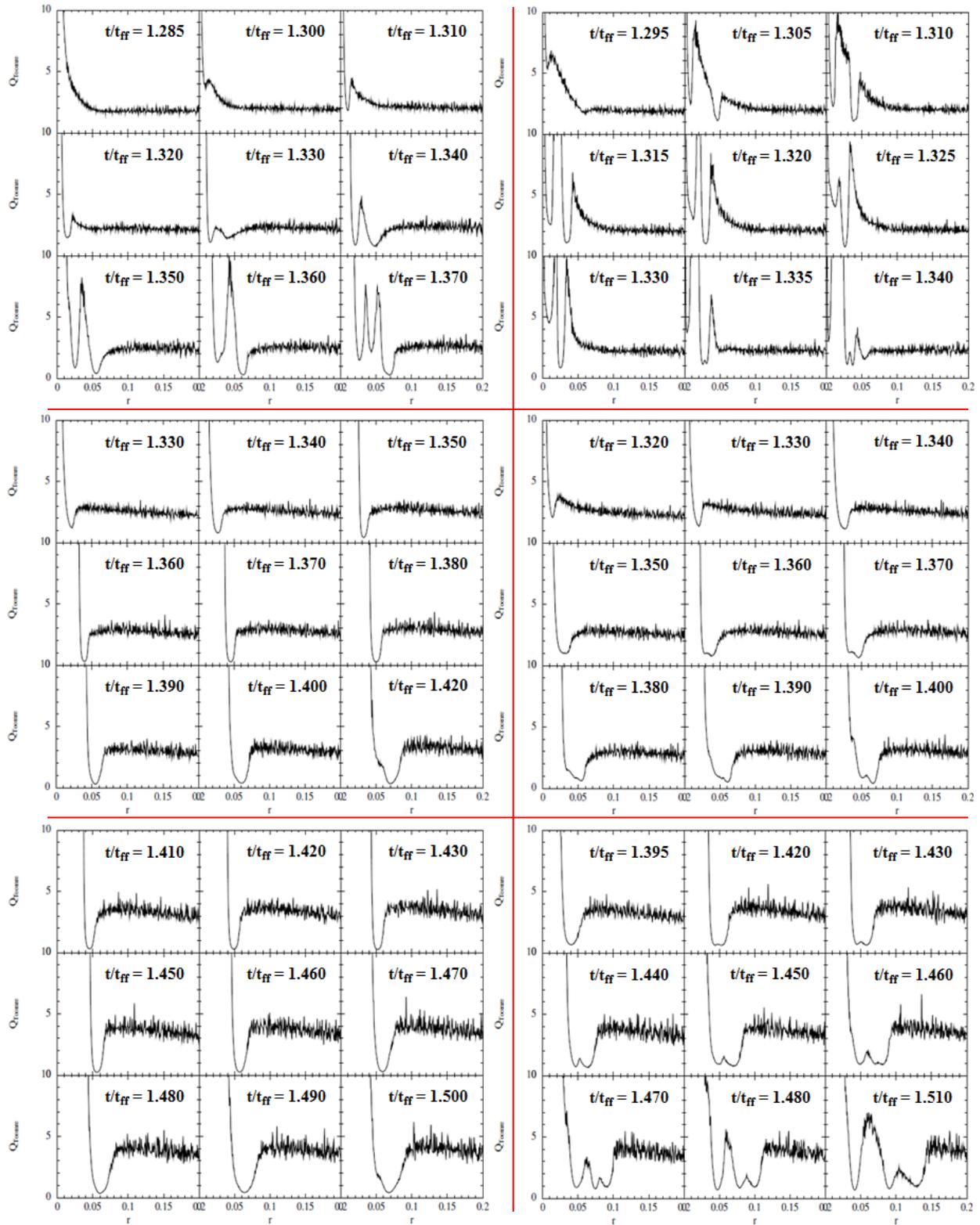

Fig.3 - From top to bottom the left and right columns represent Q-parameters values against cloud radius (r) in dimensionless units for set of models M1, M, M3 and M4, M5, M6; respectively.

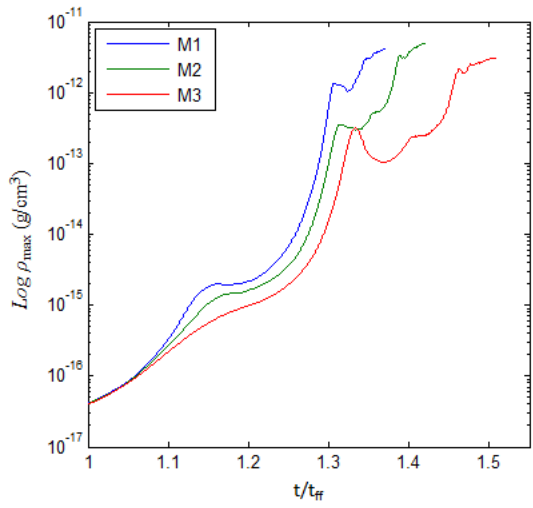 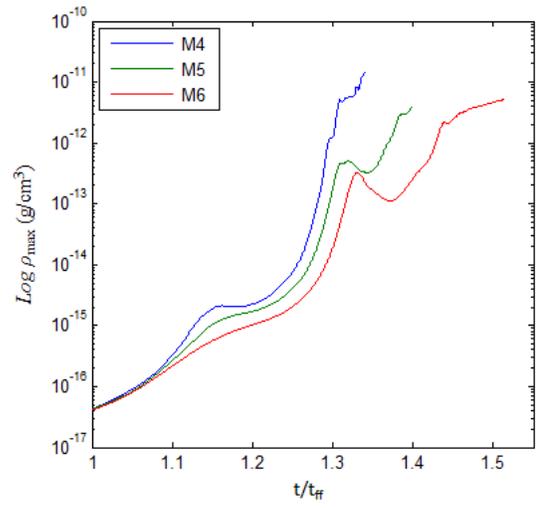

Fig.4: The time evolution of maximum density of the collapsing cloud for models M1, M2, M3 (on the left), and for models M4, M5, M6 (on the right). The free fall time of the initial cloud, $t_{ff} = 1.07 \times 10^{12}$ s (i.e. 33,968.253 yrs). Each calculation was performed with 250025 SPH particles.

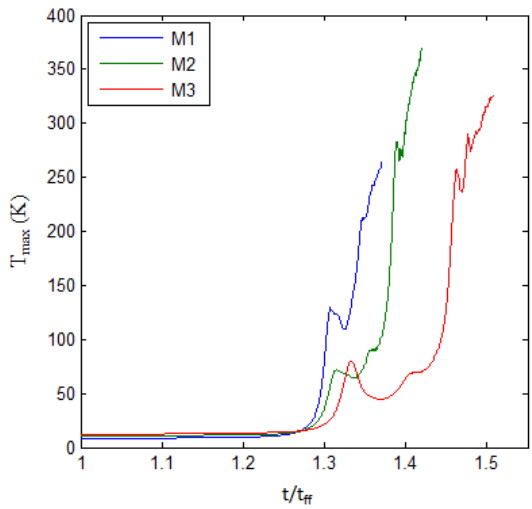 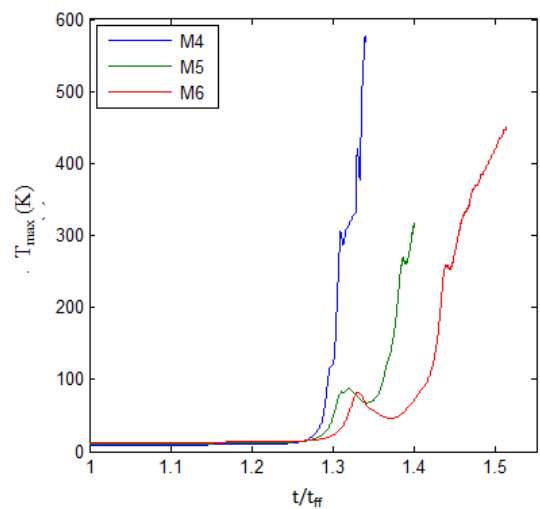

Fig.5: The time evolution of maximum temperature of the collapsing cloud for models M1, M2, M3 (on the left), & for models M4, M5, M6 (on the right). The free fall time of the initial cloud, $t_{ff} = 1.07 \times 10^{12}$ s (i.e. 33,968.253 yrs). Each calculation was performed with 250025 SPH particles.

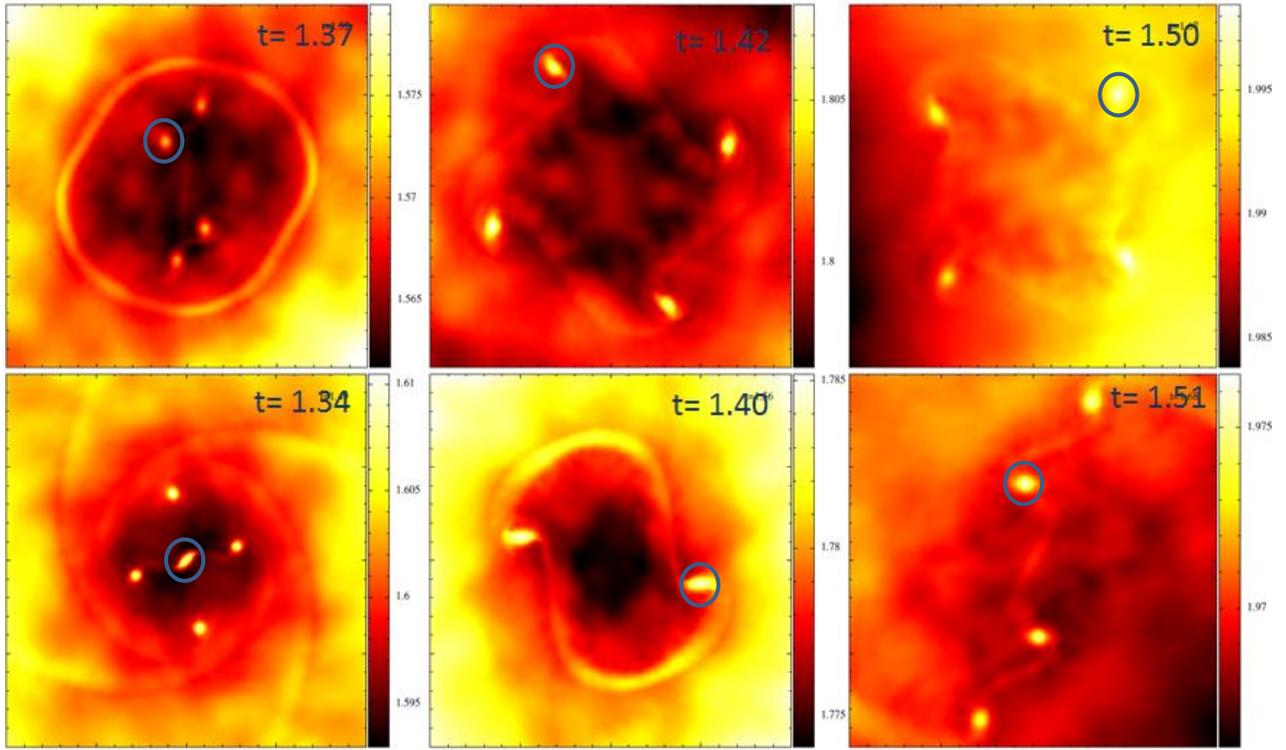

Fig.6 – Snapshots of column temperature viewed parallel to the rotation axis for the final state of evolution of cloud for models M1, M2, M3 (first row) and for models M4, M5, M6 (second row). The dimensions of each square for xy-plane follow dimensionless units as 0.1 x 0.1. Color coded bar in each panel follows log(T) in dimensionless units. Fragmentation inside a circle indicates highest thermal region. Each calculation was performed with 250025 SPH particles.

Table1: Summarizing physical parameters and respective results for models. The radius, mass, and initial density of cloud for each models remain constant as $5 \times 10^{16}$ cm, $1 M_\odot$, $3.8 \times 10^{-18}$ g/cm$^3$; respectively.

| Model | Temp$_{init.}$ T (K) | Sound speed c_iso (km/s) | Parameter (α) | Perturbation amplitude A(%) | Structure evolution |
|---|---|---|---|---|---|
| M1 | 08 | 0.148 | 0.207 | 10 | Spiral → Ring |
| M2 | 10 | 0.165 | 0.259 | 10 | Spiral → Ring |
| M3 | 12 | 0.181 | 0.311 | 10 | Ring |
| M4 | 08 | 0.148 | 0.207 | 25 | Spiral |
| M5 | 10 | 0.165 | 0.259 | 25 | Spiral → Ring |
| M6 | 12 | 0.181 | 0.311 | 25 | Ring → SGF |

Table2: Summarizing the maximum evolution time, respective maximum density and temperature attained by the cloud for collapse models.

| **Model** | M1 | M2 | M3 | M4 | M5 | M6 |
|---|---|---|---|---|---|---|
| $t_{max}$ (yrs) | 46911.303 | 48603.686 | 51602.054 | 45865.450 | 47919.127 | 51824.536 |
| $\rho_{max}$ (g/cm$^3$) | $4.2014 \times 10^{-12}$ | $5.0317 \times 10^{-12}$ | $3.1303 \times 10^{-12}$ | $1.3980 \times 10^{-11}$ | $3.9981 \times 10^{-12}$ | $5.1902 \times 10^{-12}$ |
| $T_{max}$ (K) | 263.4980 | 370.0559 | 326.8536 | 577.4594 | 318.8839 | 453.0717 |

## 4. Result & Discussion:

Ring formations at the very center of a rotating molecular cloud that may lead to fragmentation have already been reported by many authors (Leonardo Di G. Sigalotti, 1993). Growth of structure in form of spiral due to spontaneous symmetry breaking and ring because of the accretion shock in a self-gravitating molecular cloud are found in this study also dependent on initial thermal state of the system as described in Fig. 1 and Fig. 2; and summarized in table 1.

Comparing models in Fig. 1 in the context of different initial thermal states i.e. (8K, 10K, & 12K); respectively it can realized that fragmentations can be merged together only when cloud starts its collapse from an initial relatively lower temperature state such as 8K, otherwise, the thermal support of initial 10K or above may restrict fragmentations to come closer hence ruling out any possibility of merging of fragmentations during later evolution of the cloud. This trend gets even stronger when amplitude of perturbation is enhanced.

Models with higher initial thermal state are found responsible for producing ring instability which later on hosts the process of fragmentation. It is also observed that the higher initial thermal state models show a longer duration for which the ring structure remains intact before being disrupted to form individual self-gravitating fragments as can be seen in last column of Fig.1.

In the two sets of models (M1 → M3) and (M4 → M6) which can be classified with respect to the specific amplitude of azimuthal density perturbation (A) of strength 10% and 25%; respectively, it can be seen that an increase in speed of evolution in terms of attained values of maximum density and temperature can be observed as we shift from weak to strong case of perturbation. This can be seen in Fig. 4 and Fig.5; respectively which also individually infer a delay in collapse if a selected value of initial temperature is shifted from 8K to 12K. This has further summarized in table.2. This occurs mainly due to thermal support that a cloud has against the gravitational force when relatively strong initial thermal state prevails.

A comparison of models M1 and M4 in the first column of Fig. 1 has proved that the strength of initial amplitude of azimuthal density perturbation is also a key factor in defining the type of gravitational instability that is likely to appear in collapsing clouds. It has been found that weaker amplitude facilitates ring instability while a stronger results in spiral type instability. It can be argued that model M4 is not evolved as much as model M1 but considering the 7 order of magnitude rise in initial density of model M4, it can be speculated on reasonable ground that there exists hardly any chance of ring instability to come into play even at the later stage of collapse.

Considering the total mass of the cloud selected for the set of models in this study, we can expect that in the later part of evolution none of the fragments will ever acquire enough mass from the accretion reservoir and hence would eventually become a sub-stellar object as described in earlier studies of low-mass proto-star systems (Lee et al., 2009).

Spirals resulted in models of low initial temperature with strong amplitude of density perturbation are tightly wound, while those associated with weak amplitude of perturbation are relatively open. Similarly, the formation of large spiral arms is promoted in cloud that has an initial low thermal state, and the phenomenon gets stronger if relatively strong initial amplitude of perturbation is assigned.

In case of a cloud at 10K, the spiral wave pattern appears first which then later on host the ring structure within itself. As the cloud collapses further, the inner part i.e. the ring structure becomes dominant and the outer spiral train gradually fadeout leaving insignificant traces as represented in second column of Fig.1

Mass sharing among fragments at the later stages of collapse can be seen generally in all the initial thermal state models that follow initially strong amplitude of azimuthal density perturbation. However, at the final stage of model M1 (a case of initially weak amplitude of azimuthal density perturbation) the primary fragments exhibit mass sharing and an unusually appeared ring structure that surrounds the primary fragmentations also shows tendency to produce secondary fragmentation within itself as can be observed in last panel of the model M1 of Fig.1. A further investigation is indeed needed here to explore the possibility of any mass sharing between the primary and secondary fragmentations.

Secondary fragmentation also occurs in case of model M4 with a much higher temperature value as compared to other primary fragmentations. This can be seen in second row and first column of Fig. 6.

Spiral waves are the effective means of an outward angular momentum transfer during early phase of evolving disk that accumulates mass from the parent cloud. Similarly, the development of ring instability at the central region of the cloud when undergoes the process of fragmentation may also transform initial spin angular momentum of the cloud into orbital motion of self-gravitating fragments. In this context, we propose that the initial thermal state of molecular cloud is important enough to determine which mechanism of angular momentum transfer is adopted by the self-gravitating cloud as there are specific models which develop specific instability under given specific initial thermal states.

The development of spiral or ring instability has profound effects on the Q-parameter profile of the evolving disk. A sharp dip in Q-parameter values measured against radius of evolving cloud is observed and found as an indication of occurrence of spiral instability as can be seen in first row of Fig. 3 while rest of the panels show no such transient in Q-profile which in general remained as an indication of transformation from spiral to ring instability.

A rise in initial cloud temperature of even 2K from a conventional value of 10K can affect the evolution of cloud as ring instability arrives much earlier than primary fragmentation. Also as the disk stretches further, the ring density increases along with an obvious decrease in Toomre's Q-parameter values. This gives rise to the possibility of fragmentation within the ring itself.

In case of models with low initial thermal states, the gravitational instability arrives much closer to the center of the disk and a substantial outer portion of the disk remained stable against any type of instability which means that no effective heating due to gravitational instability is present in this part of the disk. This trend, however, has not been observed in higher initial thermal state models in which spiral or ring instability is experienced by the disk at much larger radii as can be analyzed in Fig.3.

Q-parameter for ring instability in Fig. 3 is also inferring the severity and the time of occurrence of centrifugal bounce experienced by the collapsing cores. No matter which amplitude of perturbation (weak/strong) is introduced, the centrifugal bounce is generally remained as a feature of molecular core having 12K as initial temperature as can be seen in last row of Fig.3. However; cores with initial temperature of 10K experience the bounce back phenomenon at the later stages of collapse when spiral instability is transformed into ring instability as indicated in second row of Fig.3. An exception, however, is observed for the case of an initially 8K core which gives birth to ring instability at the later part of evolution but only when assigned weak amplitude of azimuthal density perturbation. A quantitative analysis further infers that the collapsing cores in models M2, M3, M5 and M6 that have higher initial thermal states along with both weak and strong amplitudes of azimuthal density perturbations undergo centrifugal bounce at Q-parameter value well below unity and in some cases quite close to zero. Interestingly, when collapsing core in model M1 experiences the bounce back at a later stage of collapse then the Q-parameter value is found again fairly close to zero (see the first row last panel of Fig. 3).

Irrespective of strength of initial amplitude of azimuthal density perturbation, an increase in initial temperature of molecular cloud from 8K to 12K results in much wider disk formation (for comparison see Fig. 2).

**Acknowledgements:**

We are thankful to the referee's valuable comments and suggestions that helped to improve the manuscript. We are also thankful to the Institute of Space & Planetary Astrophysics (ISPA) for support of research projects that led us to this paper. For the computational resources we remain grateful to National University of Science & Technology (NUST) and National Centre for Physics (NCP) for giving us the access to their computational facilities.

**References:**

Arreaga-García G., Klapp J., Sigalotti, L. D. & Gabbassov, R.: 2007, Ap. J., 666, 290

Boss A.P., Harri A.T. Vanhala: 2000, Space Science Reviews, v. 92, Issue 1/2, p. 13-22

Bate and Burkert.: 1997, Monthly Notices of the Royal Astronomical Society, Volume 288, Issue 4, pp. 1060-1072.

Bally J., Langer W.D., Stark A. A., Wilson R. W.: 1987, ApJ, 312, L45

Commercon, B.,Hennebelle, P., Audit, E., Chabrier, G., Teyssier, R.: 2008, A&A Volume 482, Issue 1, pp. 371-385

Chang Won Lee, Tyler L. Bourke, Philip C. Myers, Mike Dunham, Neal Evans, Youngung Lee, Tracy Huard, Jingwen Wu, Robert Gutermuth, Mi-Ryang Kim, Hyun Woo Kang: 2009, ApJ, 693, 1290-1299

Commerçon, B.; Hennebelle, P.; Audit, E.; Chabrier, G.; Teyssier, R.:2008. Astronomy and Astrophysics, Volume 482, Issue 1, 2008, pp.371-385

Duchene, G., Bontemps, S. Bouvier, J., Andre, P., Djupvik, A. A., Ghez, A. M.: 2007, A&A, Volume 476, Issue 1, pp.229-242

Gingold & Monaghan: 1977, Monthly Notices of the Royal Astronomical Society, vol. 181, p. 375-389

Kevin L Luhman: 2012, Annual Review of Astronomy & Astrophysics, Volume 50, pp 65-106

Larson, Richard B.: 2005. Monthly Notices of the Royal Astronomical Society, Volume 359, Issue 1, pp. 211-222.

Leonardo Di G. Sigalotti: 1993, Monthly Notices of the Royal Astronomical Society, 268, pp. 625-640

Masunaga, Hirohiko; Inutsuka, Shu-ichiro.: 2000. ApJ, Volume 531, Issue 1, pp. 350-365

Myers P. C., Fuller G. A., Goodman A. A., Benson P. J.: 1991, ApJ, 376, 561

Maddalena R. J., Morris M., Moscowitz J., Thaddeus P.: 1986, ApJ, 303, 375

Price & Monaghan: 2007, Monthly Notices of the Royal Astronomical Society, Volume 374, Issue 4, pp. 1347-1358.

Price, Daniel J.: 2007, Publ. Astron. Soc. Aust., 24, 159-173

Stamatellos D., Whitworth A. P., Bisbas T., Goodwin S.: 2007, A&A, 475, 37

Tohline, J. E.: 1982. Fundamentals of Cosmic Physics, vol. 8, no. 1, p. 1-81

Toomre A.: 1964. ApJ, 139, 1217

Ungerechts H., Thaddeus P.: 1987, ApJS, 63, 645

Vanaverbeke, S., Keppens, R, Poedts, S., Boffin, H.: 2007, Computer Physics Communication, Volume 180, Issue 7, p. 1164-1182

Whitworth & Anthony P.: 2007, et al., 2007, Proceedings of the International Astronomical Union 2, IAU Symposium 237, Cambridge University Press, pp. 251-257